\documentclass{nature}
\usepackage{times}
\usepackage{xcolor}
\usepackage{graphicx}
\usepackage{amsmath}
\usepackage{physics}
\usepackage{gensymb}
\usepackage{color}
\usepackage{changes}
\usepackage{url}

\title{Physics-Guided and Physics-Explainable Recurrent Neural Network for Time Dynamics in Optical Resonances}

\author{Yingheng Tang,$^{1,2}$ Jichao Fan,$^{1}$ Xinwei Li,$^{3}$ Jianzhu Ma,$^{4}$ Minghao Qi,$^{2}$ Cunxi Yu,$^{1,\ast}$ and Weilu Gao$^{1,\ast}$}

\begin{document}

\maketitle

\begin{affiliations}
 \item Department of Electrical and Computer Engineering, University of Utah, Salt Lake City, UT 84112, USA
 \item Department of Electrical and Computer Engineering, Purdue University, West Lafayette, IN 47907, USA
 \item Division of Physics, Mathematics and Astronomy, California Institute of Technology, Pasadena, CA 91125, USA
 \item Institute of Artificial Intelligence, Peking University, Beijing, China
\end{affiliations}
\noindent $^\ast$To whom correspondence should be addressed; E-mail: cunxi.yu@utah.edu; weilu.gao@utah.edu.

\pagebreak
\begin{abstract}
Understanding the time evolution of physical systems is crucial to revealing fundamental characteristics that are hidden in frequency domain. In optical science, high-quality resonance cavities and enhanced interactions with matters are at the heart of modern quantum technologies.
However, capturing their time dynamics in real-world scenarios suffers from long data acquisition and low analysis accuracy due to slow convergence and limited time window. Here, we report a physics-guided and physics-explainable recurrent neural network to precisely forecast the time-domain response of resonance features with the shortest acquired input sequence being 7\% of full length, and to infer corresponding resonance frequencies. The model is trained in a two-step multi-fidelity framework for high-accuracy forecast, where the first step is based on a large amount of low-fidelity physical-model-generated synthetic data and second step involves a small set of high-fidelity application-oriented observational data. Through both simulations and experiments, we demonstrate that the model is universally applicable to a wide range of resonances, including dielectric metasurfaces, graphene plasmonics, and ultrastrongly coupled Landau polaritons, where our model accurately captures small signal features and learns essential physical quantities. The demonstrated machine learning algorithm offers a new way to accelerate the exploration of physical phenomena and the design of devices under resonance-enhanced light-matter interaction.
\end{abstract}

\pagebreak

Forecasting time dynamics has been a key area in many contexts of scientific research and commercial decision making, such as climate modeling\cite{MudelseeEtAl2019ER}, medical science\cite{TopolEtAl2019NM}, and finance\cite{LimEtAl2020APA}. Classical parametric models informed by domain expertise, such as autoregressive integrated moving average\cite{BoxEtAl2015}, enjoy merits of model simplicity and easy solutions while suffer from large prediction errors in strongly nonlinear and stochastic data. Machine leaning approaches provide new perspective of learning time dynamics and underlying complex representation in a data-driven manner without fixed parameters or structures\cite{GoodfellowEtAl2016}. Especially, given the impressive success of recurrent neural networks (RNNs) in natural language processing and logical interpretation of time series as sequences, various RNN architectures, such as long short-term memory and gated recurrent unit (GRU), have been employed for time series forecast applications\cite{FuEtAl2016,KongEtAl2017ITSG,NelsonEtAl2017}. However, the overfitting and local minimum problems often lead to low model efficacy\cite{HyndmanEtAl2020IJF}. Moreover, the predictive power of these models is limited to forecasting short sequences, while the high accuracy long sequence forecast is more desirable for practical scenarios\cite{ZhouEtAl2020APA2}. In addition, conventional machine learning approaches generally require significant amount of training data\cite{JiangEtAl2020NRM,MaEtAl2020NP} that is infeasible in practical context and they are also unable to extract explainable information and knowledge from the dataset. The fusion of machine learning models with empirical models is a promising effective and efficient learning philosophy to address these limitations\cite{RangapuramEtAl2018,SalinasEtAl2020IJF}, especially physics-guided machine learning methods in physical science\cite{WillardEtAl2020APA,KarniadakisEtAl2021NRP}.


High-quality optical cavities for enhancing light-matter interaction form the backbone of optical quantum technologies\cite{PellizzariEtAl1995PRL,CiracEtAl1997PRL,DuanEtAl2001N,MabuchiEtAl2002S,EnglundEtAl2007N,ObrienEtAl2009NP,Forn-DiazEtAl2019RMP}. Among them, one example showing the significance of capturing and understanding time dynamics is the terahertz time-domain spectroscopy (THz-TDS) of THz resonance structures, which continue to stimulate much interest in diverse disciplines and applications such as health\cite{SiegelEtAl2004ITMTT}, sensing and imaging\cite{TonouchiEtAl2007NP,MittlemanEtAl2018OE}, security\cite{LiuEtAl2007PI}, computing\cite{LinEtAl2018S,LiEtAl2021SR}, and communication\cite{Kleine-OstmannEtAl2011JIMTW}, in addition to quantum applications\cite{ZhangEtAl2016NP,LiEtAl2018NP}. The uniqueness of THz-TDS includes the simultaneous acquisition of amplitude and phase information, broad spectral coverage, and time-resolved capability\cite{UlbrichtEtAl2011RMP,JepsenEtAl2011LPR,NeuEtAl2018JAP}. However, the energy decaying in resonance features is notoriously slow, leading to long data acquisition time. Moreover, theoretically infinitely long THz time signal is inevitable to be practically truncated for reasonable experimental measurement and numerical simulation time or preventing undesired phenomena. Thus, scientific and technical trade-offs between data accuracy and acquisition time have to be carefully crafted. 

Here, we describe a physics-guided and physics-explainable cascaded GRU networks model to forecast long-sequence time-domain signal using short-sequence input obtained from THz-TDS numerical simulations and experiments. Instead of directly training the model using a significant amount of high-fidelity data from either computationally expensive electromagnetic solvers or time consuming experiments, we employ a two-step multi-fidelity training approach. A large number of low-fidelity physical-model-generated synthetic data is employed to first pretrain the model, which reduces the hypothesis and search space of the network for fast and efficient learning, alleviates local minimum problems for high-accuracy forecast, and generalizes model applicability. Through transfer learning using only a small set of high-fidelity application-oriented data, the pretrained model is tailored to a broad range of resonance features, including resonant dielectric metasurfaces\cite{NadellEtAl2019OE}, electro-optic graphene plasmonics\cite{JuEtAl2011NN}, and the ultrastrong coupling between electron cyclotron resonance in a high-mobility two-dimensional electron gas (2DEG) and photons in a high-quality-factor cavity\cite{LiEtAl2018NP}. The cascaded networks enable precise long sequence forecast with the shortest input time sequence being 7\% of full sequence length, suggesting the best $15\times$ data acquisition speedup. Furthermore, this model accurately captures signal features that only occupy 0.01\,\% of total signal energy in the experimentally collected data of Laudau polaritons, and simultaneously learns resonance energies of spectral features. The polariton dispersions obtained from experimentally measured and forecast time series, as well as model-learned quantities, show excellent agreement, and the derived cooperativity from forecast spectra matched experimentally reported value well\cite{LiEtAl2018NP}.

\begin{figure}
  \includegraphics[width=0.9\textwidth]{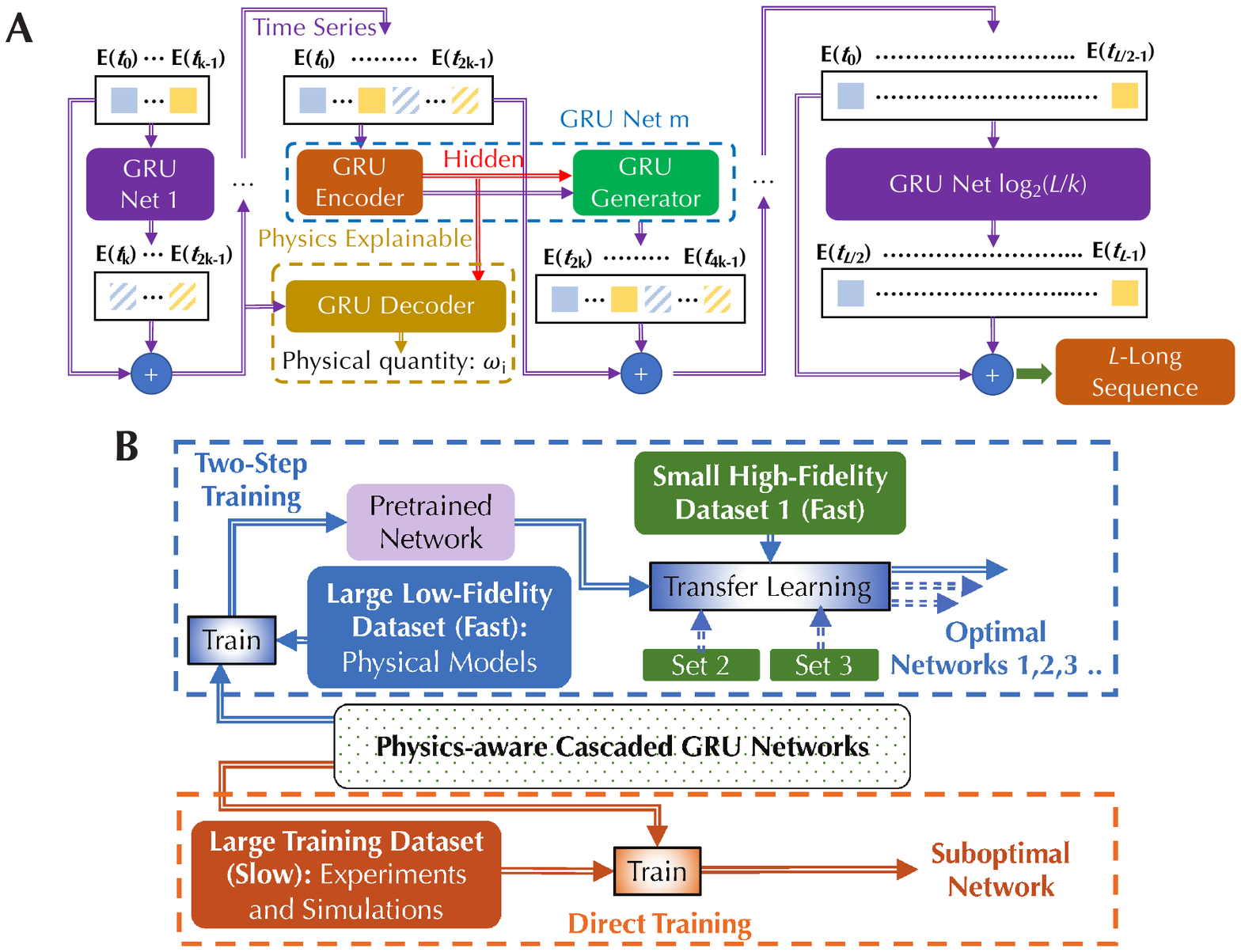}
  \caption{\label{fig:model} \textbf{Cascaded GRU networks and two-step training approach.} (a)\,Forecast of a $L-$long time signal with a $k-$long time signal using $O$(log$_2$($L/k$)) cascaded gated recurrent units (GRUs). The output sequence of the GRU at each stage is combined with the input sequence to serve as the input for the next-stage GRU. In experiments, a physics-explainable GRU network is connected to a GRU encoder in the middle stage of cascaded GRU networks that are for time series forecast. (b)\,Slow and suboptimal conventional training approach (orange path) and our fast, broadly applicable, and optimal two-step training approach (blue path). Conventional training approach needs a large number of training datasets. The generation of such large datasets requires long time of numerical simulations and experiments and sometimes is not even feasible in experimental settings. In contrast, in our multi-fidelity training approach, the model is first trained with fast generated low-fidelity synthetic data from physical models and then tailored to specific applications using only a small number of high-fidelity datasets from either numerical simulations or experiments.}
\end{figure}

Figure\,\ref{fig:model}a illustrates the model of cascaded GRU networks taking short input sequence (length $k$) and forecasting long output sequence (length $L$). 
The initial input sequence is used in the first GRU network and then combined with forecast output sequence for the input of the next-stage network. Theoretically, the minimum required number of GRU networks is $O$(log$_2$($L/k$)). In a middle stage of GRU networks, a physics-explainable GRU decoder is branched out for simultaneously learning corresponding physical quantities, resonance energies, associated with time signals (golden dashed rectangle in Fig.\,\ref{fig:model}a). Hidden states from the GRU encoder and the time series data from previous stage GRU network make the input for both physics-explainable GRU decoder and the forecast GRU generator to produce the time series output of this GRU stage. The physics-explainable GRU decoder suggests that hidden states obtained from time series forecast GRU networks bear actual physical meaning, and is also practically useful in experiments for accelerating not only the acquisition but also the explanation and understanding of time dynamics data. 

Figure\,\ref{fig:model}b displays the training process of GRU networks. Conventional training approach (orange path in Fig.\,\ref{fig:model}b) requires a large training dataset, which generally takes long time to be generated through simulations or experiments and frequently leads to suboptimal trained networks because of local minimum issues. In contrast, we utilize a physics-guided multi-fidelity two-step training approach (blue path in Fig.\,\ref{fig:model}b), where random networks are first pretrained with a large number of low-fidelity synthetic data instantaneously generated from analytical physical models based on domain expertise. The pretrained network is then fine trained through transfer learning with a small high-fidelity specific dataset from various applications, which is obtained from either numerical simulations or experiments. The fully trained network is optimal and has superior performance over the one trained through conventional approach. 

\begin{figure}
  \includegraphics[width=0.9\textwidth]{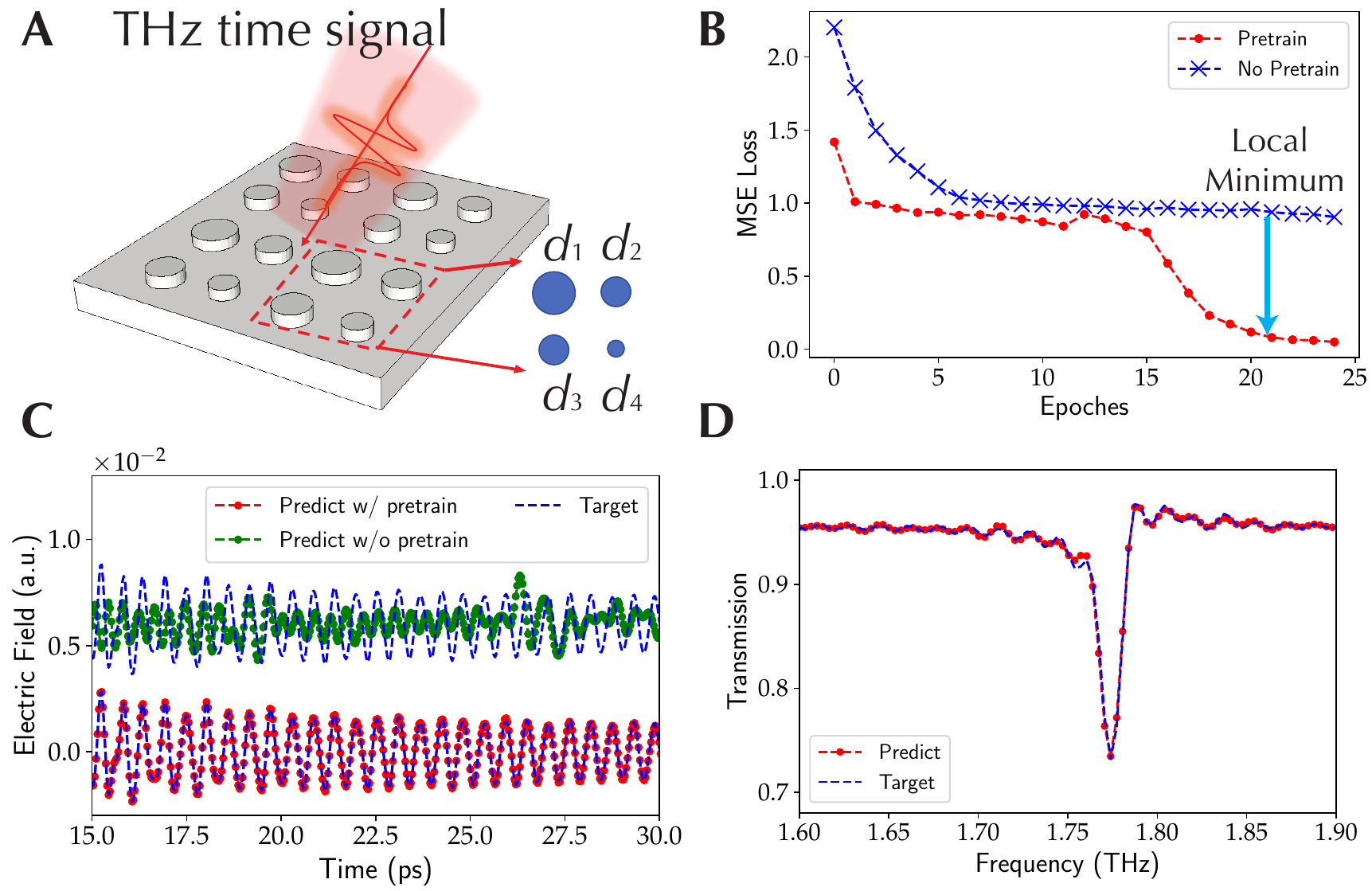}
  \caption{\label{fig:app1} \textbf{Dielectric metasurfaces.} (a)\,A generic dielectric metasurface with the unit cell consisting of four cylindrical pillars of different dimensions. (b)\,The escape from local minimum in the two-step training approach. (c)\,Forecast time signal using conventional training approach and two-step approach. The clear better forecast performance is observed in the two-step approach. (d)\,THz spectra obtained through the Fourier transformation of the full time signal and the forecast time signal using short input time sequence. Excellent agreement between forecast spectra and target spectra confirms the superior performance of our two-step approach.}
\end{figure}

We first demonstrate our model and training methodology for time series forecast in an example of resonant dielectric metasurfaces, which consists of a periodic array of unit cells with four dielectric cylindrical pillars of varying heights ($h$) and diameters ($d$)\cite{NadellEtAl2019OE}; see Fig.\,\ref{fig:app1}a. We generated training data by numerically simulating electrical field time response of different structures with randomly generated combinations of $h$ and $d$; 
detailed data generation procedure is described in \emph{Methods}. In addition to bright photonic modes, the coupling between neighboring pillars can generate sharp Fano resonances. Figure\,\ref{fig:app1}b displays the training loss as a function of training epochs using conventional and two-step approaches. Despite extensive hyper-parameter tuning in conventional training approaches (see more details on hyper-parameters in \emph{Methods}), we clearly observe local minimum issue that is prevailing in non-convex optimization problems and leads to poor forecast (blue dashed line with cross markers). 

In order to solve this problem and improve forecast accuracy, we employ a multi-fidelity training framework\cite{WillardEtAl2020APA}. Most physical resonance features and structures render themselves as a sum of damped oscillations in time signals, which follow a general mathematical form $\Sigma_i A_ie^{-\alpha_i t}\text{sin}\\(\omega_i t + \phi_i)$, where $A_i$ is the amplitude factor, $e^{-\alpha_i t}$ describes harmonics envelope decay, $\text{sin}(\omega_i t + \phi_i)$ is an oscillating carrier with a resonance frequency $\omega_i$ and initial phase $\phi_i$, and $i$ enumerates all resonance features. We employed this analytical model to generate low-fidelity synthetic data, by randomly choosing $\omega_i$ within the frequency of interest, and arbitrarily selecting $\alpha_i$, $\phi$ and the number of resonances; detailed generation procedures, data pre-processing, and exemplary synthetic data are in \emph{Supplementary Information Sections 1, 2, and Fig.\,S1}. All GRU networks were pretrained with these synthetic data and then fine trained via transfer learning with a small set of high-fidelity time-domain data of the dielectric metasurface obtained through full-wave finite domain time difference (FDTD) simulations (see \emph{Methods} for training data generation). The ratio of the length of final sequence over that of input sequence, $L/k$, is $\sim15$. As clearly indicated in Figs.\,\ref{fig:app1}b and c, Mean-Squared-Error (MSE) and the accuracy of forecast time series utilizing the multi-fidelity training approach outperforms those obtained using conventional approach. Furthermore, in the frequency domain, the spectra that is calculated through the Fourier transformation of forecast time signals obtained using the model trained by the multi-fidelity approach also show excellent agreement with the target spectra calculated by running full-time simulations. 

\begin{figure}
  \includegraphics[width=0.9\textwidth]{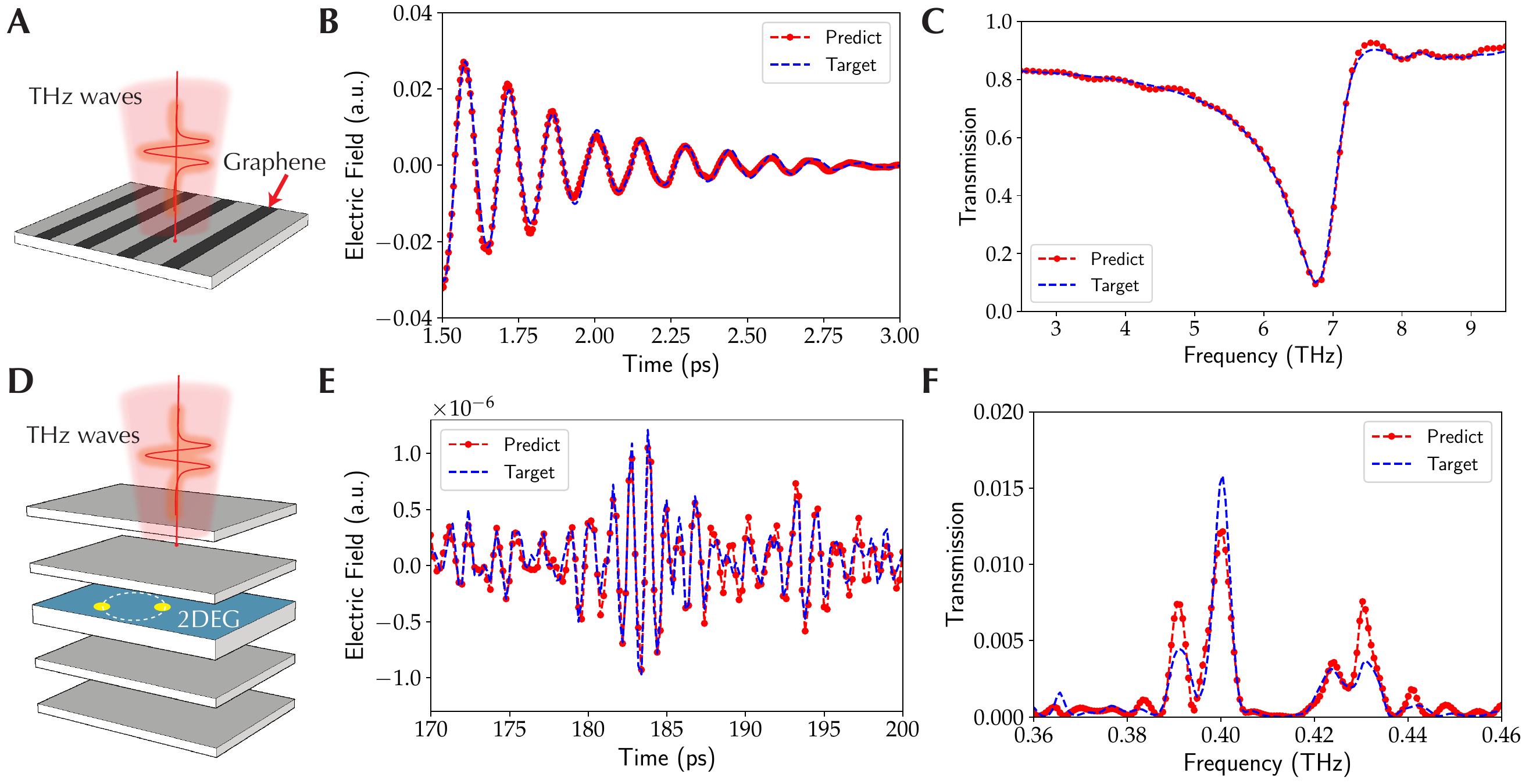}
  \caption{\label{fig:trans} \textbf{Model transfer to graphene plasmonics and Laudau-polaritons.} (a) -- (c)\,Graphene plasmonics. (a)\,Schematics of graphene ribbons supporting localized plasmonic resonance. (b)\,Time domain and (c)\,corresponding frequency domain response for predicted time signal and full target time signal, respectively. (d) -- (f)\,Laudau-polaritons in strongly coupled photons in an one-dimensional photonic crystal cavity with the electron cyclotron resonance in high-mobility two-dimensional electron gases (2DEG). (e)\,Time domain and (f)\,corresponding frequency domain response under specific magnetic field for predicted time signal and full target time signal, respectively.}
\end{figure}

The transfer learning technique greatly generalizes the applicability of pretrained networks. The similar damped oscillation signature of the time-domain signals in most resonance features physically guarantees the feasibility of such generalization. We demonstrated the approach generalization to two physically distinct resonance features, active graphene plasmonics\cite{JuEtAl2011NN} and ultrastrongly coupled Laudau polaritons\cite{ZhangEtAl2016NP,LiEtAl2018NP}, in addition to dielectric metasurfaces. Specifically, periodically patterned monolayer graphene ribbons, which are expected to support localized plasmonic THz resonance from bounded carriers by ribbon boundaries, are simulated to obtain time-domain signals; see Fig.\,\ref{fig:trans}a. The ribbon width and graphene Fermi level are arbitrarily selected; details can be found in \emph{Methods}.  With similar pretrained networks and a small set of short input sequences ($L/k\sim4$), both time and frequency domain forecast show excellent agreement with the target time signal (Fig.\,\ref{fig:trans}b) and spectra (Fig.\,\ref{fig:trans}c) obtained from FDTD simulations. 

Moreover, an ultrahigh-mobility 2DEG inside a high-quality-factor one-dimensional THz photonic cavity under magnetic fields displayed ultra-narrow Landau polaritons\cite{LiEtAl2018NP}; see Fig.\,\ref{fig:trans}d. Their spectra were experimentally measured with THz-TDS under high magnetic fields. There are in total 71 spectra under various magnetic fields from 0\,T to 4.5\,T, and we used 24 spectra as training data to fine-train synthetic-data-trained model and the rest 47 spectra as test data. These 24 spectra were chosen as ones taken under magnetic fields either uniformly or randomly distributed between 0\,T and 4.5\,T. We then used the model trained by selected 24 measurements to forecast the rest 47 time signals under other magnetic fields with corresponding short input sequence to reveal full dispersion of Laudau polaritons.
Compared with the other two simulation examples, experimental data are much more noisy and the number of dataset is very limited. Moreover, our Laudau polariton features locate inside the defect mode of the stopband of photonic crystals, and the signal energy of interest only occupies 0.01\,\% of total signal energy; see Fig.\,S2 in \emph{Supplementary Information}. All these factors make the forecast using conventional direct training approach nearly impossible in practice. The multi-fidelity two-step approach instead significantly improves the predictive power and accuracy, and thus small but important resonance features are captured. The loss function is also the MSE loss calculated from predicted and target time signals; details on the loss function selection are in \emph{Supplementary Information Section 3 and Fig.\,S3}. Figures\,\ref{fig:trans}e and f display one representative time signal and corresponding spectra from Fourier transformation both aligning well with target experimental results. 

For Laudau polariton experiments, in addition to cascaded GRU networks for time-domain signal forecast, we added a physics-explainable GRU network as shown in Fig.\,\ref{fig:model}a in the middle (2nd) GRU network to infer resonance energies associated with time signals. The cascaded GRU networks are first trained for time series forecast, and then the physics-explainable GRU network is trained with the hidden states and intermediate sequences from time series forecast GRU networks as input; see \emph{Methods} for detailed training process. After training, the short time signals from test data are input into the full model, and both forecast time signals and corresponding resonance energies are simultaneously generated. 

\begin{figure}
  \includegraphics[width=0.9\textwidth]{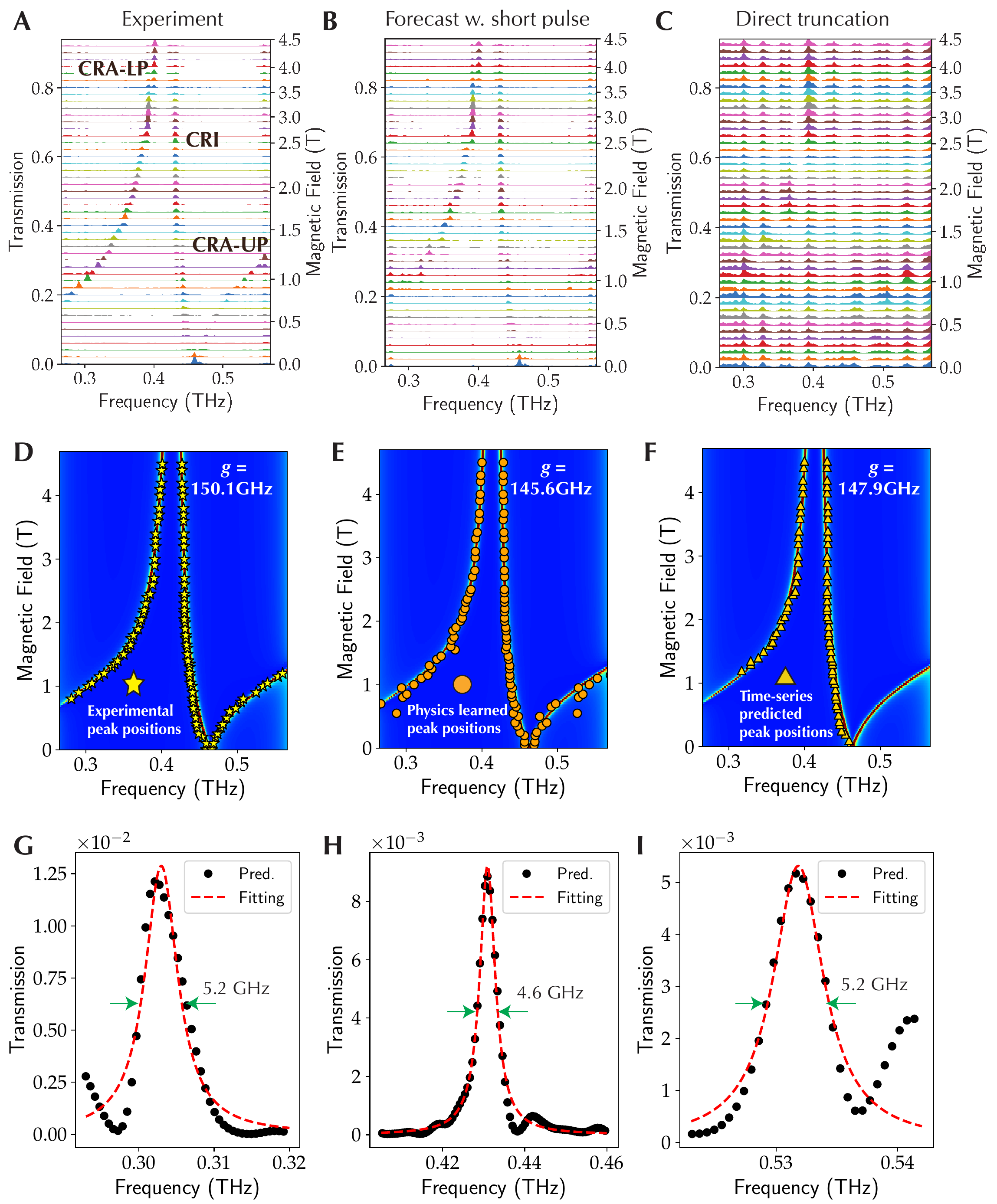}
  \caption{\label{fig:exp} \textbf{Experimental verification of models in Laudau-polaritons.} Linearly polarized THz transmittance spectra under various magnetic fields obtained through the Fourier transformation of (a)\,experimentally acquired time signals, (b)\,predicted time signals with a much shorter input experimental time sequence, and (c) short input experimental time signals with zero padding. CRA-LP and CRI branches clearly observed in experimental spectra are reproduced in predicted spectra, where the CRA-UP branch is more noisy. The spectra obtained from zero-padded input time signal is completely random. Simulated transmittance colour contour map using transfer matrix method to fit (d)\,experimental spectra shown in (a), (e)\,model-learned resonance frequencies, and (e)\,predicted spectra shown in (b). Yellow stars in (d), orange circles in (e), and gold triangles in (f) mark the peak positions extracted from experimental spectra, learned from physics-explainable GRU network, and extracted from predicted spectra. The extracted coupling rates $g$ are 150.1\,GHz, 145.6\,GHz, and 147.9\,GHz, respectively. Lorentzian fits (red dashed line) of predicted spectra for (g)\,the UP peak at zero detuning (1T) (h)\,the LP peak at zero detuning, and (i)\,the CRI mode at 3T. The obtained full-width at half-maximum values are indicated by green arrows. }
\end{figure}

Figures\,\ref{fig:exp}a and b display all 47 experimental and predicted spectra obtained from the Fourier transformation of corresponding time signals, where $L/k$ is $\sim6$ for predicted time sequences. All essential physical features expected in linearly polarized transmission spectra are well reproduced, including cyclotron-resonance-active lower polaritons (CRA-LPs), CRA upper polaritons (CRA-UPs), and CR-inactive (CRI) modes\cite{LiEtAl2018NP}; see \emph{Supplementary Information Section 4 and Figs.\,S4 and S5} for more data in both time and frequency domains. In a stark contrast, the spectra obtained directly from short input sequences with appropriate zero padding, as shown in Fig.\,\ref{fig:exp}c, display completely random patterns with all features missing. The distinct difference between Figs.\,\ref{fig:exp}a, \ref{fig:exp}b, and \ref{fig:exp}c highlights the necessity of long data acquisition to capture essential THz features if no prediction is employed, as well as the high predictive power and accuracy of our model of cascaded GRU networks and training approach. Note that the results do not depend on how 24 training spectra are selected and the average validation loss displays small variance when training spectra are randomly shuffled for different magnetic fields; see \emph{Fig.\,S6 and Supplementary Information Section 4}. 

Furthermore, yellow stars in Fig.\,\ref{fig:exp}d, orange circles in Fig.\,\ref{fig:exp}e, and gold triangles in Fig.\,\ref{fig:exp}f show extracted peak positions from experimental measured spectra that are the Fourier transformation of time signals, model-learned resonance frequencies, and extracted peaks from predicted spectra, respectively. We utilized transfer matrix formalism to calculate transmission spectra for samples under various magnetic fields, through which the coupling ratio $g$ can be extracted; details can be found in \emph{Supplementary Information Section 5} and Ref.\,\cite{LiEtAl2018NP}. Figs\,\ref{fig:exp}d, e, and f also display calculated transmittance colour contour map to match extracted and learned peaks. Despite noisy CRA-UP branch in predicted spectra, the fitting can be done uniquely. The extracted coupling ratios, $g_\text{exp}$, $g_\text{explain}$, and $g_\text{pred}$ from experimental spectra, physics-explainable GRU decoder, and predicted spectra, are 150.1\,GHz, 145.6\,GHz and 147.9\,GHz. This excellent agreement confirms not only the precise forecast of cascaded GRU networks but also generated hidden states bear actual and explainable physical meanings, which are resonance energies. 

With obtained predicted dispersion, the magnetic field corresponding to the conditions at zero detuning and far away from zero detuning can be determined. As shown in Figs.\,\ref{fig:exp}g, h, and i, we predicted spectra of CRI mode at 3\,T and the CRA-UP and CRA-LP peaks at zero detuning using short input sequence with $L/k\sim2$ using 24 training spectra and determined their spectral line widths to be 4.6\,GHz, 5.2\,GHz and 5.2\,GHz, respectively. Thus, the cooperativity from predicted sequence is 3663, which is very close to the experimental value 3513\cite{LiEtAl2018NP}. In addition to the acceleration of experimental data acquisition for mapping out Laudau polariton dispersion and determining corresponding physical quantities, our model is useful to assist and guide experiments by dynamically forecasting time signals through active learning. We utilized measured spectra from 4.5\,T to a magnetic field close to zero detuning (e.g., $B_\text{m} = 1.4$\,T) as training spectra to forecast the spectrum under next lower magnetic field ($B_\text{m-1}$). We then augmented the training spectra by adding the measured spectra under $B_\text{m-1}$ for forecasting the spectra under $B_\text{m-2}$, and this process keeps moving to the lowest magnetic field. The anomalous increase of training loss during this active learning process indicates new time signal and spectral features, such as additional transmission peaks in the stopband around zero detuning, and helps experimentalists to make their decisions; see \emph{Fig.\,S7 and Supplementary Information Section 4} for more details. 


Our physics-guided and physics-explainable cascaded GRU networks combine complementary advantages of both physics-based models and machine learning approaches for fast, efficient, and accurate forecast of long time signals in optical resonances with a minimum of 7\,\% input sequence, which corresponds to a best $15\times$ speedup of data acquisition. The multi-fidelity two-step training framework enables the model to be generalized to varieties of contexts of resonance features and structures because of the similarity of their physical appearance. This model is especially efficient in practical experimental scenarios, where long data acquisition is inevitable and the collection of a large number of data is almost infeasible. The incorporation of synthetic data in training process together with transfer learning technique significantly reduces the required number of training data to 24 spectra for forecasting experimental spectra of Laudau polaritons, and improves the predictive power and accuracy for capturing signal features that only occupy 0.01\,\% of total signal energy. This approach is promising to accelerate the discovery of new phenomena in complex systems through the analogy with more accessible optical system\cite{LiEtAl2018S} and the exploration of device functionalities under resonant light-matter interactions.

\newpage
\begin{methods}

\subsection{Training data generation for dielectric metasurface and graphene plasmonics.}

Finite domain time difference (FDTD) time-domain simulations implemented in commercial Ansys Lumerical software are used for the training data generation in the examples of the dielectric metasurface and graphene plasmonics. For the dielectric metasurface example, we used four silicon cylindrical rods as the base structure with periodic boundary condition. Different data were generated by randomly selecting the radii of four cylindrical rods. The radius is chosen in the range from $39.5\,\mu$m to $44.5\,\mu$m with a step resolution of $0.25\,\mu$m. We generated a total number of $5500$ samples and using $5000$ of them as the training set and the rest $500$ as the validation set. For the graphene plasmonics example, graphene monolayer is modeled as a 2D rectangle conducting sheet in Lumerical material library, including both interband and intraband contributions. Fermi level and scattering rate are two parameters used to calculate dielectric constants used for the software. The dataset is generated by randomly sweeping graphene ribbon width and chemical potential. The width range is between $3.8\,\mu$m to $13.8\,\mu$m with a step of $0.2\,\mu$m. The chemical potential ranges from $0.18$\,eV to $0.41$\,eV with a step of $0.01$\,eV. We generated a total number of $1000$ samples and using $800$ of them as the training set and the rest $200$ as the validation set. 

\subsection{Training of recurrent neural networks with GRU.} 

The multi-fidelity two-step training contains two steps: pretrain and fine-train (transfer learning). During the pretrain stage, the synthetic data are divided into pieces with different sequence length and fed into the model. During the fine-train process, the actual data (simulations and experimental measurements) are used to further update all parameters of the model. The GRU network we used here is a four-layer structure with a size of $200$ hidden states. In the pretrain stage, batch size is set as $128$ and the total epoch number is $80$. The \texttt{Adam} optimizer is used with the initial learning rate set at $5\times10^{-4}$. The learning rate decays every 30 epochs with the decay rate 0.01. The detailed hyper-parameters are summarized in the \emph{Supplementary Information Section 6 Table 1}.

In the example of Landau polaritons, once the model for time dynamics forecast is fully trained, a physics-explainable GRU decoder is connected to the trained encoder of the 2nd GRU to receive the encoded hidden states and processed time signal outputs. We trained this physics-explainable GRU network with resonance frequency labels associated with time-domain signals, while we kept weights of the rest cascaded GRU networks unchanged and only updated this branch GRU decoder. The decoder is a four-hidden layer GRU network similar to the decoders in time signal forecasting networks. In limited 71 experimental spectral data, we utilized 24 data for training and 47 data for testing. This physics-explainable network was also trained using the \texttt{Adam} optimizer with a learning rate of $5\times10^{-5}$. The total epoch for the training is 300 with the batch size of 1. The learning rate decays every 100 epochs with the decay rate 0.1.

\end{methods}

\pagebreak
\begin{addendum}
\item [Data availability] Upon publication, source data will be provided with this paper. All other data that support the plots within this paper and other findings of this study will be available from the corresponding authors upon reasonable request. 

\item [Code availability] Upon publication, the codes relevant to neural network models and associated simulation data will be publicly available on \emph{GitHub}. The experimental data will be available from the corresponding authors upon reasonable request. All other codes that support the plots within this paper and other findings of this study are available from the corresponding authors upon reasonable request.

\item J.\,F. and W.\,G. thank the support from the University of Utah start-up fund. C.\,Y. thanks the support from grants NSF-2019336 and NSF-2008144. 

\item [Author contributions] C.\,Y. and W.\,G. conceived the idea and designed the project. Y.\,T. performed the modeling and calculations with the help of J.\,F., J.\,M., M.\,Q., C.\,Y. and W.\,G. X.\,L. helped with the analysis of Landau polariton data. Y.\,T. and W.\,G. wrote the manuscript. All authors discussed the manuscripts and provided the feedback. 

\item[Competing Interests] The authors declare that they have no competing financial interests.

\item[Correspondence] Correspondence and requests for materials should be addressed to Cunxi Yu (email: cunxi.yu@utah.edu) and Weilu Gao (email: weilu.gao@utah.edu).
\end{addendum}

\newpage

\newpage

\newpage

\newpage

\newpage
\bibliography{weilu.bib} 

\end{document}